\documentclass[prl,twocolumn,preprintnumbers,amsmath,amssymb,superscriptaddress]{revtex4}
\usepackage{epsfig}

\usepackage{graphicx}
\usepackage{dcolumn}
\usepackage{bm}
\usepackage{float}
\usepackage{hyphenat}
\usepackage[dvipsnames]{xcolor}
\usepackage{physics}
\usepackage[T1]{fontenc}
\usepackage{gensymb}
\usepackage[utf8]{inputenc}
\usepackage{amsmath}
\usepackage{amsthm}
\usepackage{ulem}

\DeclareUnicodeCharacter{2212}{\ensuremath{-}}

\newcommand{\heriotwatt}{Institute of Photonics and Quantum Sciences, SUPA, Heriot-Watt University, Edinburgh EH14 4AS, UK}
\newcommand{\TsukubaKenji}{Research Center for Electronic and Optical Materials, National Institute for Materials Science, 1-1 Namiki, Tsukuba 305-0044, Japan}
\newcommand{\TsukubaTakashi}{Research Center for Materials Nanoarchitectonics, National Institute for Materials Science,  1-1 Namiki, Tsukuba 305-0044, Japan}
\newcommand{\insa}{Université de Toulouse, INSA-CNRS-UPS, LPCNO, 135 Avenue de Rangueil, 31077 Toulouse, France}

\begin{document}

\title{Quadrupolar and Dipolar Excitons in Bilayer 2$H$-MoSe$_2$}

\author{Shun Feng}
\affiliation{\heriotwatt}

\author{Aidan  J. Campbell}
\affiliation{\heriotwatt}

\author{Bibi Mary Francis}
\affiliation{\heriotwatt}
 
\author{Hyeonjun Baek}
\affiliation{Department of Physics, Sogang University, Seoul, South Korea}
\author{Takashi Taniguchi}
\affiliation{\TsukubaTakashi}
\author{Kenji Watanabe}
\affiliation{\TsukubaKenji}
 
\author{Iann C. Gerber}
\affiliation{\insa}
 \author{Brian D. Gerardot}
 \email{B.D.Gerardot@hw.ac.uk}
 \affiliation{\heriotwatt}

\author{Mauro Brotons-Gisbert}
\email{m.brotons_i_gisbert@hw.ac.uk}
\affiliation{\heriotwatt} 

\date{\today}

\begin{abstract}
We report the experimental observation of quadrupolar exciton states in the reflectance contrast spectrum of 2$H$-stacked bilayer MoSe$_2$. The application of a vertical electric field results in a quadratic energy redshift of these quadrupolar excitons, in contrast to the linear energy splitting observed in the coexisting dipolar excitons within the bilayer MoSe$_2$. We perform helicity-resolved reflectance contrast measurements to investigate the spin and valley configurations of the quadrupolar exciton states as a function of applied vertical electric and magnetic fields. Comparing our results with a phenomenological coupled-oscillator model indicates that the electric- and magnetic-field dependence of the quadrupolar exciton states can be attributed to the intravalley and intervalley hybridization of spin-triplet interlayer excitons with opposite permanent dipole moments, mediated by interlayer hole tunneling. These results position naturally stacked MoSe$_2$ bilayers as a promising platform to explore electric-field-tunable many-body exciton phenomena.

\end{abstract}

\maketitle

A rich tapestry of excitons – tightly bound electron-hole pair quasiparticles – arise in atomically thin transition metal dichalcogenide semiconductors (TMDs) and their heterostructures. The optical response of monolayer TMDs exhibits a Rydberg series of intralayer excitons which form at the $\pm$K points of the Brillouin zone and possess unique spin and valley selection rules \cite{xiao2012coupled,mak2012control,chernikov2014exciton,stier2018magnetooptics}. Stacking two different monolayer TMDs gives rise to the formation of interlayer excitons (IXs), in which the electron and hole wavefunctions are localized on different layers \cite{rivera2015observation,nayak2017probing} and are highly tunable with an applied electric field \cite{jauregui2019electrical,ciarrocchi2019polarization,baek2020highly,tang2021tuning,shimazaki2020strongly}. The interlayer excitons and their properties are highly sensitive to the specific atomic arrangement between the adjacent layers, which can yield intrinsic ferroelectric fields \cite{sung2020broken,liang2022optically} or periodic moir{\'e} potentials with selection rules determined by the atomic registry \cite{seyler2019signatures,brotons2020spin}. Adding more layers to the TMD homo- or hetero-structures can result in interlayer excitons with even larger static dipole moments (with a wavefunction spread across more than two layers) \cite{arora2017interlayer,zhang2023every,lian2023exciton,feng2024highly}. Recently, a new species of interlayer excitons with a characteristic quadratic energy response on applied vertical electric fields has been observed in the photoluminescence emission  of the so-called quadrupolar excitonic resonances in symmetric TMD-based heterotrilayers \cite{lian2023quadrupolar, yu2023observation, li2023quadrupolar, du2024new}. These quadrupolar excitons have been experimentally \cite{lian2023quadrupolar, yu2023observation, li2023quadrupolar} and theoretically \cite{deilmann2024quadrupolar} shown to originate from the hybridization of the valence \cite{lian2023quadrupolar, yu2023observation, du2024new} and conduction \cite{li2023quadrupolar} bands in the topmost and bottommost layers of the symmetric heterotrilayers. Beyond the attractive “simple” picture of an interlayer exciton being formed by carriers highly localized on their respective layers, there is growing evidence which suggests that interlayer excitons (especially in homobilayers) are composed of an admixture of different exciton wave functions and this mixture can be tuned with applied electric fields \cite{gerber2019interlayer,leisgang2020giant,peimyoo2021electrical,sponfeldner2021capacitively,feng2024highly}, providing a natural platform to explore tunable many-body exciton complexes.

In this Letter, we report the experimental observation of quadrupolar exciton (QX) states in the reflectance contrast spectrum of naturally occurring 2$H$-stacked bilayer MoSe$_2$. Unlike previously reported dipolar excitons in this system, which exhibit a linear energy shift with an out-of-plane electric field \cite{kipczak2023analogy,feng2024highly}, these QX states show a quadratic redshift, suggesting a distinct origin underlying their quadrupolar nature. We employ a phenomenological coupled-oscillator model to unravel both the exciton species and the hybridization mechanism responsible for the formation of the quadrupolar exciton states. We attribute the origin of these excitonic resonances to the intravalley and intervalley hybridization of spin-triplet interlayer excitons and observe a very good agreement with the experimental electric-field-dependent evolution of the QXs energy. Further, we carry out helicity-resolved reflectance contrast measurements to investigate the magneto-optical properties of the QXs as a function of applied vertical electric and magnetic fields. Our results reveal vanishing exciton Land{\'e} g-factors for the hybrid QX states and a degree of circular polarization that can be tuned by the applied electric field. Both observations are captured by our phenomenological model and are compatible with the proposed picture of coupled spin-triplet interlayer excitons as the origin of the QX states.

\begin{figure}
    \begin{center}
    	\includegraphics[scale= 0.45]{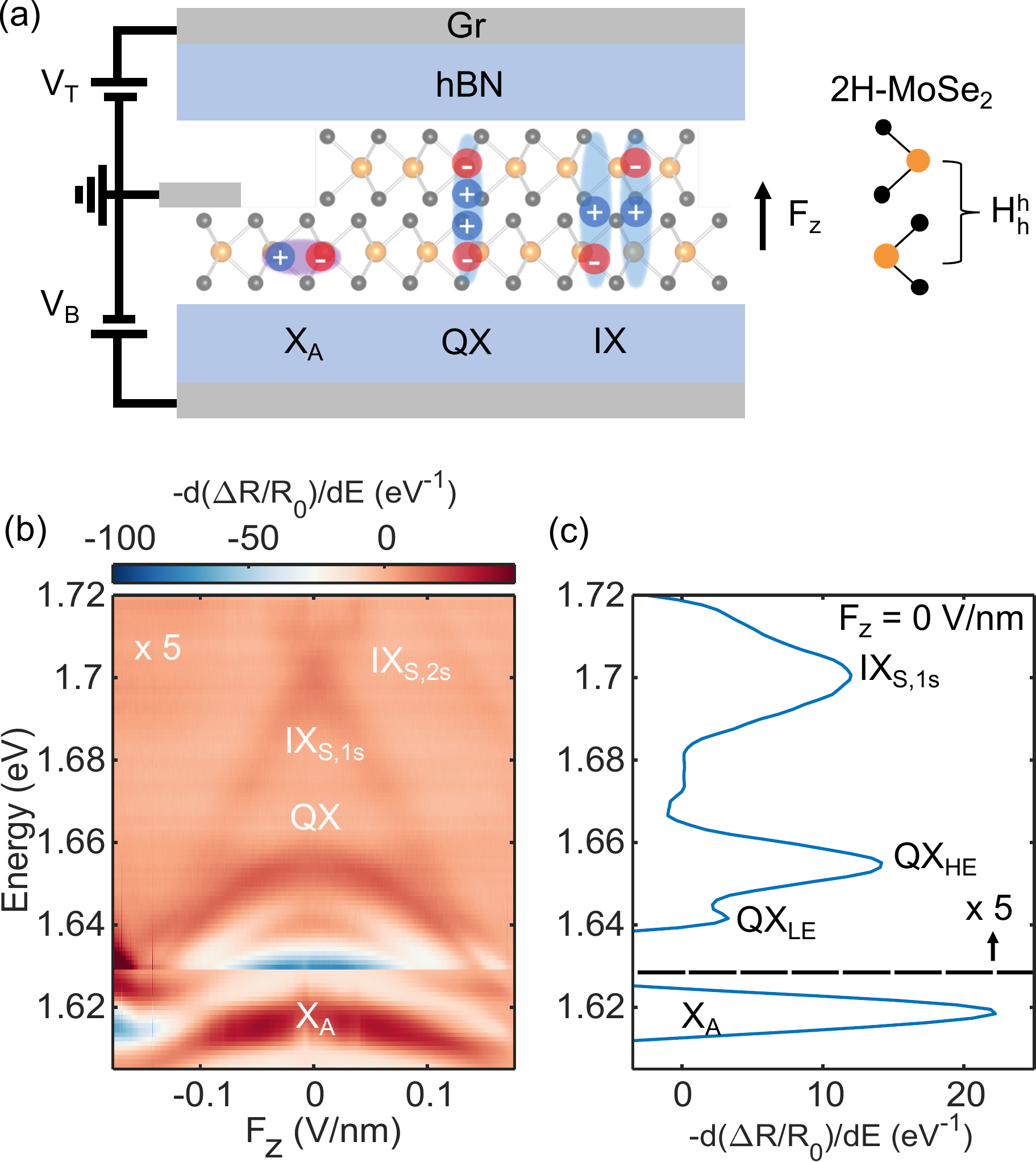}
    \end{center}
    \caption{Dipolar and quadrupolar excitons in 2L 2H-MoSe$_2$. (a) Schematics of the dual-gated 2H-MoSe$_2$ device structure. The orange and gray dots represent  Mo and Se atoms, respectively. Electrons (red circles) at the $\pm$K-valleys, localized either in the bottom or top MoSe$_2$ layer, are strongly bound to holes (blue circles) delocalized across the two layers, creating different species of excitons: intralayer ($X_A$), quadrupolar ($QX$), and interlayer ($IX$) excitons. An out-of-plane electric field with magnitude $F_z$ can be applied. (b) Electric field dependence of the first derivative of the reflectance contrast spectra with respect to photon energy (d($\Delta$R/R$_0$)/dE) in 2L 2H-MoSe$_2$ in the spectral range 1.605 - 1.72 eV. The experimental d($\Delta$R/R$_0$)/dE values for energies above $\sim$1.63 eV have been multiplied by a factor 5 to enhance the exciton visualization. (c) Single d($\Delta$R/R$_0$)/dE spectrum at $F_z=0$ V/nm.}
    \label{fig1}
\end{figure}

Our experiments employ a natural bilayer MoSe$_2$ flake with 2$H$ stacking, which belongs to the symmetry group $D_{3d}$ and retains inversion symmetry. The bilayer is incorporated into a dual-gated device geometry [Fig. \ref{fig1}(a)]. The 2$H$-MoSe$_2$ flake is encapsulated by hexagonal boron nitride (h-BN) layers of nearly identical thickness ($d\sim$ 18 nm). Few-layer graphene flakes serve as electrical contacts for the MoSe$_2$ crystal as well as for the top and bottom gates [see Supplemental Materials for more details on the device fabrication]. In our experiments, the MoSe$_2$ contact is grounded, while gate voltages of magnitudes V$_T$ and V$_B = -V_T$ are applied to the top and bottom gates, respectively. This configuration allows us to apply an out-of-plane electric field ($F_z$) with a magnitude $F_z =\frac{{V_{B}\cdot\epsilon_{BN}}}{d\cdot{\epsilon_{TMD}}}$ (where $\epsilon_{BN}$ ($\epsilon_{TMD}$) is the dielectric permittivity of h-BN (TMD)),while maintaining the carrier concentration in the MoSe$_2$ sample close to charge neutrality (see details in the End Matter section).

We perform reflection contrast ($\Delta$R/$R_0$) spectroscopy at cryogenic temperature (4 K) as a function of $F_z$ at charge neutrality, where $\Delta$R = $R_s$ − $R_0$, and $R_s$ ($R_0$) is the intensity of the light reflected by the flake (substrate). Figure \ref{fig1}(b) shows the contour plot of the energy derivative of the reflectance contrast spectrum d($\Delta$R/R$_0$)/dE as a function of the applied electric field $F_z$ in the spectral region $\sim$1.605 - 1.72 eV. The experimental d($\Delta$R/R$_0$)/dE values for energies above $\sim$1.63 eV have been multiplied by a factor 5 to enhance the exciton visualization. Figure \ref{fig1}(c) shows the d($\Delta$R/R$_0$)/dE spectrum at $F_z=0$ V/nm (V$_T=0$ V). At $F_z=0$, we observe four excitonic resonances in this spectral range. The strongest exciton resonance at low energy ($\sim$1.615 eV) corresponds to the ground state of the intralayer $A$ exciton in bilayer 2$H$-MoSe$_2$ (X$_A$) \cite{kipczak2023analogy,feng2024highly}. The dimmer resonance at $\sim$1.7 eV corresponds to the spin-conserving (spin-singlet) optical transition from the ground state interlayer exciton in bilayer 2$H$-MoSe$_2$ (IX$_{S,1s}$) \cite{kipczak2023analogy,feng2024highly}. As previously observed in Ref. \cite{feng2024highly}, Fig. \ref{fig1}(b) shows that the application of a vertical electric field $F_z$ results in the DC-Stark-induced linear energy splitting of IX$_{S,1s}$ states with opposite permanent electric dipoles \cite{feng2024highly}. We note that optical transitions from the first excited Rydberg state of IX$_{S,1s}$ (IX$_{S,2s}$) are also visible in Fig. \ref{fig1}(b) at larger $F_z$-field values, exhibiting a linear DC Stark shift with the same dependence on the applied $F_z$ \cite{feng2024highly}. At $F_z=0$, two additional exciton resonances, separated only by $\sim$16 meV, are also observed in the energy range between X$_A$ and IX$_{S,1s}$ in both Figs. \ref{fig1}(b) and \ref{fig1}(c). Strikingly, the application of a vertical electric field results in a quadratic energy redshift of this excitonic doublet [see Fig. \ref{fig1}(b)], which contrasts with the behaviors observed for X$_A$ and IX$_{S,1s}$. We note that the presence and the quadratic energy shift of this excitonic doublet are reproduced in different spatial positions of our sample [see Supplemental Material for results in additional sample positions]. Similar quadratic dependence on the electric field has recently been observed in the photoluminescence emission energy of quadrupolar excitonic resonances in symmetric TMD-based heterotrilayers \cite{lian2023quadrupolar, yu2023observation, li2023quadrupolar, du2024new}. However, the observation of excitons with a quadrupolar nature in the reflectance contrast spectrum of a natural TMD bilayer is intriguing.

We tentatively attribute the origin of these excitonic resonances to quadrupolar excitons arising from the coupling of spin-triplet interlayer excitons with opposite static out-of-plane electric dipoles, in which a hole delocalized across the bilayer is bound to layer-localized electrons in the top or bottom layers. Therefore, we label the high- and low-energy hybrid excitonic peaks in the doublet as QX$_{HE}$ and QX$_{LE}$, respectively. Note that we discard the possibility of electron wavefunction delocalization as the origin of the QXs states, as interlayer electron hopping at the $K$ point of the Brillouin zone has been shown to be forbidden in 2$H$ TMD homobilayers due to the $C_3$ symmetry of the $d_{z^2}$ orbitals of the conduction band states at $\pm K$ \cite{gong2013magnetoelectric,pisoni2019absence,gerber2019interlayer}, whereas interlayer hole delocalization has been theoretically predicted and experimentally observed in several bilayer TMDs with 2$H$ stacking \cite{gerber2019interlayer,paradisanos2020controlling,leisgang2020giant,lorchat2021excitons,feng2024highly}. 

The hybrid nature of the QX states near $F_z = 0$ also has direct consequences on their spatial structure. Figure \ref{fig1}(a) shows a cartoon of the expected wavefunction spatial configurations of X$_A$, IX$_{S,1s}$ and QX at $F_z=0$ (V$_T=0$ V). The coupling of interlayer excitons with opposite permanent dipoles (i.e., involving electrons localized in the top or bottom layers) results in a symmetric QX wavefunction in the out-of-plane direction. Such a symmetric spatial arrangement of the carriers results in an effective vanishing static electric dipole, which agrees well with the negligible Stark shift observed for QXs at low $F_z$. At finite $F_z$, the hole charge distribution becomes asymmetric, resulting in an increasing electric dipole moment and a transition from quadratic to linear energy shift (i.e., a transition from a quadrupolar to a dipolar interlayer exciton nature), as observed in Fig. \ref{fig1}(b). Moreover, the interlayer nature of QX$_{LE}$ and QX$_{HE}$ is further supported by the reduced oscillator strength of these optical transitions, which we estimate to be $\sim20$ times smaller than that of X$_A$ [see Supplemental Material Figs. S3(a)-(d)]. However, the coexistence of interlayer excitons in bilayer 2$H$-MoSe$_2$ with both dipolar and quadrupolar nature raises the question about the spin-layer configurations of the coupled interlayer excitons forming the QXs as well as their hybridization mechanism.

\begin{figure}
    	\begin{center}
    		\includegraphics[scale= 0.49]{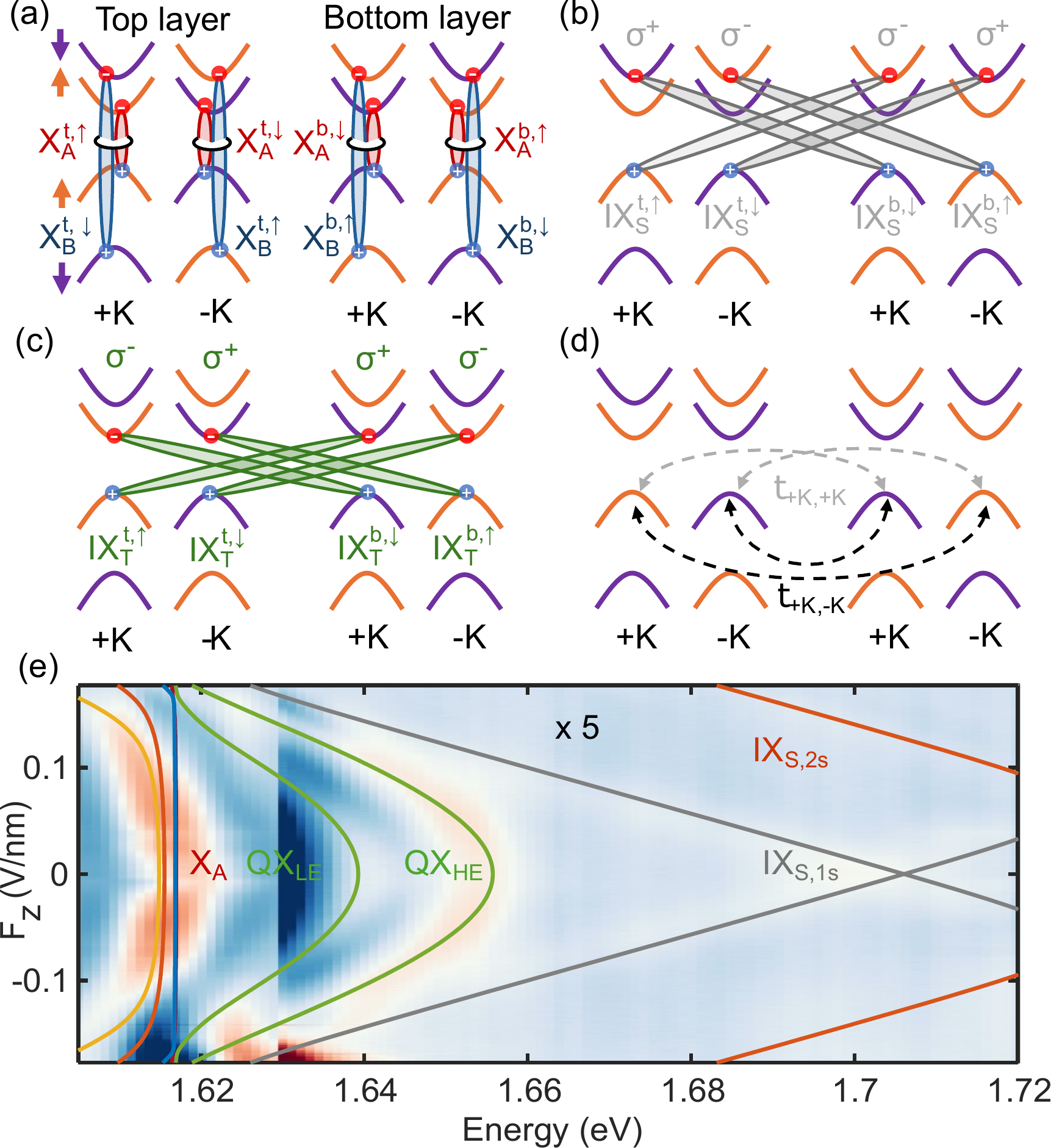}
    	\end{center}
        \caption{Spin, valley, and layer configurations of the ground states of the main momentum-direct exciton resonances in bilayer 2H-MoSe$_2$: (a) A and B intralayer excitons. (b) Spin-singlet interlayer excitons. (c) Spin-triplet interlayer excitons and their expected optical selection rules. The superscripts $t/b$ and $\uparrow/\downarrow$ denote the layer (top/bottom) and electron spin configuration (up/down) of the hole forming the exciton, respectively. (d)  Main interlayer hole tunneling processes at $\pm$K included in our phenomenological model to account for exciton hybridization. (e) Comparison of the experimental $F_z$ dependence of the first derivative of the reflectance contrast spectra with respect to photon energy d($\Delta$R/R$_0$)/dE in 2L 2H-MoSe$_2$ and the energies calculated with our phenomenological model.}
        \label{fig2}
\end{figure}

Interestingly, we observe avoided energy crossings between the QX branches with both X$_A$ and IX$_{S,1s}$ at $|F_z|>0$. Such avoided crossings provide additional insight into the hybrid nature of QXs in our bilayer 2$H$-MoSe$_2$ sample. We employ a phenomenological model in which the hybridization between the different excitons is treated as a coupling between oscillators with resonance energies corresponding to the excitonic transitions at $F_z = 0$. In our model, we take into account the spin, valley, and layer degrees of freedom of each exciton species. Figures \ref{fig2}(a)-(c) show the spin, valley, and layer configurations of the ground states of the different momentum-direct exciton resonances included in the model. The superscripts $t/b$ and $\uparrow/\downarrow$ denote the layer (top/bottom) and electronic spin configuration (up/down) of the hole forming the exciton, respectively. Figure \ref{fig2}(a) shows a sketch of the spin-layer configuration of intralayer X$_A$ excitons as well as their intravalley mixing with B excitons \cite{guo2019exchange, sponfeldner2021capacitively,feng2024highly}. Figure \ref{fig2}(b) shows the spin-layer configuration of spin-conserving (spin-singlet) interlayer IX$_{S,1s}$ states in bilayer 2$H$-MoSe$_2$ \cite{kipczak2023analogy,feng2024highly}, which are responsible for the presence of dipolar excitons and their characteristic linear energy splitting observed in Fig. \ref{fig1}(b) \cite{feng2024highly}. In addition to spin-singlet interlayer exciton states, the quadratic energy redshift as a function of $F_z$ observed for QX$_{LE}$ and QX$_{HE}$ in Fig. \ref{fig1}(b) suggests the presence of interlayer excitons with a different spin-valley configuration, which we attribute to a spin triplet configuration (IX$_{T,1s}$). Figure \ref{fig2}(c) shows a sketch of the spin-layer configuration for IX$_{T,1s}$ in bilayer 2$H$-MoSe$_2$, in which the holes in the topmost valence band of one MoSe$_2$ layer bound to electrons in the lowest spin-orbit-split conduction band of the other MoSe$_2$ layer in the same $\pm K$ valley. Although such spin-flip optical transitions are normally forbidden in monolayer TMDs \cite{wang2018colloquium}, they become symmetry allowed in bilayer 2H TMDs due to the combination of a hidden local spin polarization within each monolayer and the absence of horizontal mirror plane symmetry \cite{yu2018brightened,gilardoni2021symmetry}.

\begin{figure*}
    \begin{center}
    \includegraphics[scale= 0.35]{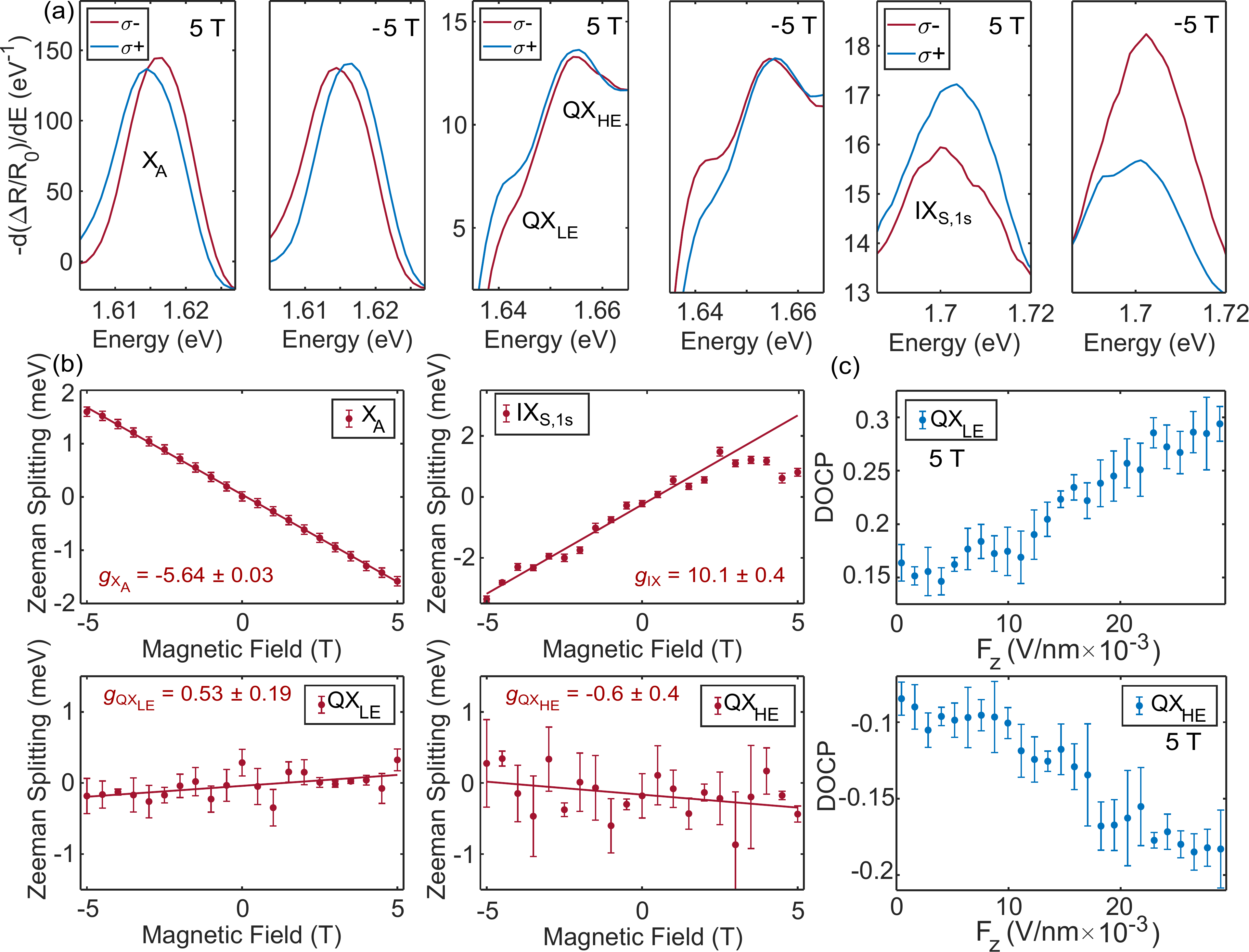}
    \end{center}
    \caption{Magneto-optical properties of excitons in bilayer 2H-MoSe$_2$. (a) $\sigma^+$- (blue) and $\sigma^-$-polarized (red) d($\Delta$R/R$_0$)/dE spectra in the energy ranges of X$_A$ (left), QXs (middle), and IX$_{S,1s}$ (right) for applied magnetic fields $B=\pm 5$ T. (b) Experimental $\Delta E(B)$ (red dots) for magnetic fields ranging from -5 T to 5 T for X$_A$ (top left),  IX$_{S,1s}$ (top right), QX$_{LE}$ (bottom left) and QX$_{HE}$ (bottom right). The red solid lines represent linear fits from which we extract the exciton g-factors. (c) Degree-of-circular polarization (DOCP) of QX$_{LE}$ (top) and QX$_{HE}$ (bottom) as function of $F_z$ at an applied magnetic field of 5 T.  
    }
    \label{fig3}
\end{figure*}

The hybridization of the different excitonic transitions in our model is included as a phenomenological exciton coupling mediated by interlayer hole tunneling at $\pm K$ [see Fig. \ref{fig2}(d)]. The intravalley and intervalley hole tunneling between the topmost valence bands in the different layers results in the hybridization of intralayer and interlayer excitons, as well as in intravalley and intervalley coupling of IX$_T$ states with opposite permanent electric dipoles [see Supplemental Materials for detailed exciton couplings]. Note that the intravalley hole tunneling between the topmost valence bands in different MoSe$_2$ layers is enabled by the intralayer A and B exciton admixture [see Fig. \ref{fig2}(a)], as reported for both bilayer MoS$_2$ and bilayer MoSe$_2$ \cite{sponfeldner2021capacitively,feng2024highly,kipczak2023analogy,slobodeniuk2019fine}. In our calculations, the energies of the bare exciton states at $F_z=0$ and the slope of the DC Stark shift for IX$_{S,1s}$ and IX$_{S,2s}$ are set to match the corresponding experimental values, while the values of the coupling strengths between the different exciton states are left as free parameters that can be tuned to match our experimental data and are independent of $F_z$ [see Supplemental Material for more details]. We also assume that the decoupled (bare) IX$_{T,1s}$ have the same Stark shift as IX$_{S,1s}$. Figure \ref{fig2}(e) shows the calculated energies of the resulting hybrid exciton states as a function of $F_z$ superimposed on the experimental data. The phenomenological model captures well the presence of QX$_{LE}$ and QX$_{HE}$, their quadratic energy redshift, as well as the avoided energy crossings with both X$_A$ and IX$_{S,1s}$ with increasing $F_z$, further supporting our hypothesis about their origin. It is worth noting that the model also predicts the presence of two additional quadrupolar excitonic transitions at higher energies with a $F_z$-induced behavior symmetric to that of QX$_{HE}$ and QX$_{LE}$ [see Supplemental Material]. These higher energy quadrupolar exciton transitions are not observed in our experimental data. This can be understood from the low oscillator strength expected for their anti-bonding (antisymmetric) character as compared to the bonding character of the QX states \cite{li2023quadrupolar,lian2023quadrupolar,deilmann2024quadrupolar}. 

To gain further insight into the origin of the QX states in bilayer 2$H$-MoSe$_2$, we perform helicity-resolved reflectance contrast measurements under out-of-plane magnetic fields at $F_z=0$. Figure \ref{fig3}(a) shows $\sigma ^+$- (blue) and $\sigma ^-$-polarized (red) d($\Delta$R/R$_0$)/dE spectra in the energy ranges of X$_A$ (left), QXs (middle), and IX$_{S,1s}$ (right) for applied magnetic fields $B=\pm 5$ T. The application of a vertical magnetic field $B$ results in the Zeeman shift of the optical transitions of each exciton state at $\pm K$, leading to energies $E^{\sigma^{\pm}}(B)=E_0^{\sigma^{\pm}}\pm1/2 g\mu_0B$, where $E^{\sigma ^{\pm}}$ is the energy of the $\sigma ^{\pm}$-polarized transition, $E_0^{\sigma^{\pm}}$ represents the transition energy at $B=0$, $g$ is the exciton Land{\'e} g factor, and $\mu_0$ is the Bohr magneton. We focus first on the already known X$_A$ and IX$_{S,1s}$ states \cite{kipczak2023analogy,feng2024highly}. The magnitude and sign of the Zeeman splitting can be estimated experimentally as $\Delta E(B)=E^{\sigma ^+}-E^{\sigma ^-}$. The red dots in the top panels of Fig. \ref{fig3}(b) show the experimental $\Delta E(B)$ for magnetic fields ranging from -5 T to 5 T for X$_A$ (left) and IX$_{S,1s}$ (right). The red solid lines represent linear fits from which we extract g-factors of $-5.64\pm0.03$ and $10.1\pm0.4$ for X$_A$ and IX$_{S,1s}$, respectively. These values are in very good agreement with previously reported g-factors for both excitonic species \cite{kipczak2023analogy,feng2024highly}. It is worth noting that the estimated g-factor for IX$_{S,1s}$ agrees very well with the experimental values (ranging between 8.6 and 9.7 \cite{kipczak2023analogy,feng2024highly}) as well as the theoretical values obtained by density functional theory calculations (8.7 \cite{kipczak2023analogy}) for the spin-singlet (spin-conserving) interlayer exciton in bilayer 2$H$-MoSe$_2$, further corroborating our IX$_{S,1s}$ assignment. More importantly, we note that the sign and magnitude of the exciton g-factor for both $X_A$ and IX$_{S,1s}$ can be effectively described by a more simplistic “atomic picture”, in which the g-factor of the bands hosting the electron-hole pairs are assumed to be equal to the sum of their spin, orbital, and valley magnetic moment contributions \cite{aivazian2015magnetic}. Within this model, we estimate a g-factor of -13.7 for the bare (uncoupled) IX$_{T}$ states [see Supplemental material for detailed calculation]. However, our phenomenological model predicts that at $F_z = 0$ both QX$_{LE}$ and QX$_{HE}$ originate from an equal mixture of the four IX$_{T}$ states represented in Fig. \ref{fig2}(c), which renders an effective Land{\'e} g-factor $g\approx0$ for the QX states [see Supplemental Material]. This prediction agrees well with the small g-factors obtained from the experimental Zeeman shifts of the QXs states [see bottom panels of Fig. \ref{fig3}(b)], which are consistent across different spatial positions in the sample [see Supplemental Material for magneto-optical results in additional sample locations].

Although the intravalley and intervalley coupling of IX$_{T}$ states prevents experimental access to the g-factor of the bare IX$_{T}$ states, the degree of circular polarization (DOCP) of QX$_{LE}$ and QX$_{HE}$ contains useful information about the QX states. The application of a vertical electric field at a given applied magnetic field leads to a progressive increase of the weight of a different IX$_{T}$ state in the nature of QX$_{LE}$ (IX$^{\uparrow,b}_T$) and QX$_{HE}$ (IX$^{\downarrow,b}_T$) [see Supplemental Material]. Based on the selection rules expected for IX$_T$ in a bilayer TMD system with 2$H$-stacking, these optical transitions should emit photons with $\sigma^+$ (IX$^{\uparrow,b}_T$) and $\sigma^-$ (IX$^{\downarrow,b}_T$) polarization \cite{yu2018brightened}, which should result in a progressive increase and decrease of the DOCP of QX$_{LE}$ and QX$_{HE}$, respectively. Based on the selection rules expected for IX$_T$ in a bilayer TMD system with 2H-stacking, these optical transitions should emit photons with $\sigma^+$ (IX$^{\uparrow,b}_T$) and $\sigma^−$ (IX$^{\downarrow,b}_T$) polarization (see Supplemental Material) \cite{yu2018brightened}, which should result in a progressive increase and decrease of the DOCP of QX$_{LE}$ and QX$_{HE}$, respectively. This prediction agrees well with our experimental results in Fig. \ref{fig3}(c) in the $F_z$ range where the QXs do not hybridize strongly with other excitonic transitions (i.e., $0\leq F_z\leq0.029 $ V/nm). We note that the qualitative behavior of the DOCP is not a particular feature of the spin-triplet configuration (see Supplemental Material), but a further corroboration that the QX states originate from the intravalley and intervalley hybridization of IXs in bilayer MoSe$_2$ with atomic interlayer registry $H_h^h$.

Finally, we employ the $GW$+BSE method to obtain an atomic-scale picture of the possible two-particle electron-hole transitions in the energy range between X$_A$ and IX$_S$ [see Supplemental Material for computational details]. We analyze the calculated spectra with respect to the applied electric field and find that the only possible single excitations present, which are direct in reciprocal space, correspond to spin-forbidden X$_A$-type transitions. All together, our $GW$+BSE results show that the properties of the QX states cannot be described by a two-particle picture, which further supports our many-body picture of hybridized IX$_T$ states as the origin of the QX transitions. Unfortunately, our current BSE scheme implementation does not allows us to account for transitions beyond the two-particle picture.

In summary, we report the experimental observation of quadrupolar and dipolar exciton states in the reflectance contrast spectrum of 2$H$-stacked bilayer MoSe$_2$. We find that both the electric- and the magnetic-field dependence of the QXs states can be attributed to the intravalley and intervalley hybridization of spin-triplet IXs mediated by interlayer hole tunneling. Further, our magneto-optical measurements discard the possibility of spin-singlet IXs as the origin of the QX resonances in bilayer MoSe$_2$. However, the reason behind the stronger exciton-exciton coupling between spin-triplet IXs compared to the spin-singlet IX species remains an open question. We speculate that this difference might be related to the electronic states involved in spin-singlet and spin-triplet IXs, which makes spin-triplet IXs share both electron and hole states with the intralayer A exciton in a four-particle-QX picture [see Supplemental Material]. Our results promote naturally stacked MoSe$_2$ bilayers as a promising platform to investigate electrically controlled many-body exciton interactions.

\textit{Note added:} After submission of our manuscript, a related work on quadrupolar excitons in bilayer 2$H$-MoSe$_2$ has been published \cite{jasinski2025quadrupolar}.

\textit{Acknowledgments} - We thank Thierry Amand for fruitful discussions. This work was supported by the EPSRC (grant nos. EP/P029892/1, EP/L015110/1, and EP/Y026284/1). S.F. was supported by a Marie Skłodowska-Curie Individual Fellowship H2020-MSCA-IF-2020 SingExTr (No. 101031596). M.B.-G. is supported by a Royal Society University Research Fellowship. B.D.G. is supported by a Chair in Emerging Technology from the Royal Academy of Engineering. K.W. and T.T. acknowledge the support from the JSPS KAKENHI (Grant Numbers 21H05233 and 23H02052) , the CREST (JPMJCR24A5), JST and World Premier International Research Center Initiative (WPI), MEXT, Japan. I.C.G acknowledges the CALMIP initiative for the generous allocation of computational time, through Project No. p0812, as well as GENCI-CINES, GENCI-IDRIS, GENCI-CCRT for Grant No. A016096649.

\bibliography{Manuscript}

\end{document}